\documentclass[prc,twocolumn,showpacs,showkeys]{revtex4}
\usepackage{epsf}
\usepackage{wrapfig}
\usepackage{amsmath}
\def\g{\gamma}
\def\m{\mu^2}
\def\mk{m_k^2}
\def\mf{m_{f_0}}
\def\ms{m_{\sigma}}
\def\me{m_{\eta}^2}
\def\mpi{m_{\pi^{\pm}}}
\def\ep{\epsilon}
\def\G{\Gamma}
\def\i{\prime}
\def\am{(\alpha_1-\beta_1)}
\def\ap{(\alpha_1+\beta_1)}
\def\amn{\am_{\pi^0}}
\def\amc{\am_{\pi^{\pm}}}

\def\dcn{\Delta(\am)}
\def\gg0{\g \g\to \pi^0 \pi^0}
\def\s{\sigma}
\def\ggc{\g\g\to\pi^+\pi^-}

\def\ggn{\g\g\to\pi\pi}

\def\mpp{M_{++}}
\def\mm{M_{+-}}
\def\sp{s'}
\def\tp{t'}

\def\fp{f_0\to\pi\pi}
\def\fk{f_0\to kk}

\def\gfp{\G_{\fp}}
\def\gfk{\G_{\fk}}
\def\gf0{\G_{0f_0}}

\def\unit{10^{-4} {\rm fm}^3}
\def\be{\begin{equation}}
\def\ee{\end{equation}}
\def\beq{\begin{eqnarray}}
\def\eeq{\end{eqnarray}}

\begin{document}
\title{Relation between the dipole polarizabilities of charged and neutral 
pions}
\vskip 0.5cm

\author{L.V.~Fil'kov}
\email[E-mail: ]{filkov@sci.lebedev.ru}
\affiliation{Lebedev Physical Institute, Leninsky Prospect 53,
Moscow 119991, Russia}
\vskip 1cm

\begin{abstract}

Using the fact that the contribution of the states with isospin $I=0$ in the 
difference 
of the amplitudes of the processes $\ggc$ and $\gg0$ is very 
small, we have analyzed the dispersion sum rules for the difference
between the dipole polarizabilities
of the charged and neutral pions  as a function of
the $\s$ meson parameters. Then taken into account the current perturbation
value of $\amn=-1.9$, we have found $\amc=9.4 \div 8.2$ for values of
the $\s$ meson parameter within the region: $\ms=400 \div 550$ MeV, 
$\G_{\s}=400\div 600$, $\G_{\s\to\g\g}=0 \div 3$ keV. 
It has been shown that the value of the decay width of the $h_1(1170)$ meson 
into $\g\pi^0$ can be found if the difference $\amc$ is reliably determined
from the experiment. Estimation of
the optimal value of the decay width $\s\to\g\g$ has given
$\G_{\s\to\g\g}\lesssim 0.7$keV.

\end{abstract}
\vskip 0.3cm


\maketitle

\section{Introduction}

Pion polarizabilities are the fundamental structure parameters
characterizing the behavior of the pion in an external electromagnetic
field. Dipole polarizabilities arise as ${\cal O}(\nu_1 \nu_2)$ terms in 
the expansion of the non-Born amplitudes of Compton scattering in powers 
of the initial and final photon energies $\nu_1$ and $\nu_2$.
In terms of the electric $\alpha_1$ and magnetic
$\beta_1$ dipole polarizabilities, the corresponding effective
interaction has the form:
\begin{equation}
H^{(2)}_{eff}=-\frac12\,4\pi\,(\alpha_1\,\vec{E}^2+\beta_1\,\vec{H}^2).
\end{equation}

The dipole polarizabilities 
measure the response of the hadron to quasistatic electric and magnetic
fields. In what follows, these parameters are given in units $\unit$. 

The values of the pion polarizabilities are very sensitive to the predictions 
of different theoretical models. Therefore, an accurate experimental 
determination of them is very important for testing the validity of such 
models.

At present,
the value of the difference of the charged pion dipole polarizabilities
\def\amc{\am_{\pi^{\pm}}} found from radiative $\pi^+$ meson 
photoproduction from protons \cite{mainz} is equal to
$11.6\pm 1.5_{stat}\pm 3.0_{syst}\pm 0.5_{mod}$ and close to the value
obtained from scattering of high energy $\pi^-$ mesons
off the Coulomb field of heavy nuclei in Serpukhov \cite{antip} 
and equal to $13.6\pm 2.8\pm 2.4$.
On the other hand, these values differ from the prediction of the
chiral perturbation theory (ChPT) (($4.7\div 6.7$) \cite{gasser}).
The experiment of the Lebedev Physical Institute on radiative pion
photoproduction from protons \cite{lebed} has given
$\alpha_{1\pi^+}=20\pm 12$.
This value has large error bars but nevertheless shows a large discrepancy 
with regard to the ChPT predictions, as well.

The preliminary result of the COMPASS collaboration 
$\amc=5.0\pm 3.4_{stat}\pm 1.2_{syst}$ has been found by studying the
$\pi^-$ meson scattering off the Coulomb field of heavy nuclei \cite{compass}. 
This result is more close to the ChPT prediction. However to obtain this 
result the authors used very big values of momentum transfer 
$Q^2_{max}\approx 5\times 10^{-3}$(Gev/c)$^2$. In this region an interference between
the Coulomb and nuclear amplitudes should be taken into account 
\cite{fil4,fil-PS,walch}. 
It should be noted that the authors of  work \cite{antip}
chose $Q^2<6\times 10^{-4}$ (GeV/c)$^2$ to guarantee that the contribution of
the strong interaction below the Coulomb peak is negligible.   

The charged pion polarizabilities can be also found by studying  
the process $\ggc$.    
Investigation of the process $\ggn$ at low and middle energies was carried
out in the framework of different theoretical models 
and, in particular, in the frame of dispersion relations (DR).
In Ref. \cite{fil1,fil2,fil3} we have analyzed the processes $\gg0$ and
$\ggc$ using DRs with subtractions for the invariant amplitudes $\mpp$ and
$\mm$ without partial-wave expansions. The subtraction constants
have been uniquely determined in these works through the pion polarizabilities.
The values of the polarizabilities have been found from the fit of the 
experimental data of the processes $\ggc$ and $\gg0$ up to 2500 MeV and
2250 MeV, respectively. As a result, we have found
$\amc=13.0^{+2.6}_{1.9}$ and  $\amn=-1.6\pm 2.2$. The result for $\amc$ is
in good agreement with the values obtained in Ref.\cite{mainz,antip,lebed}
whereas it is at variance with the ChPT prediction. 
 
In the works \cite{bab,holst,kal,garcia} 
the dipole polarizabilities of charged pions have been determined
from the experimental data of the process $\ggc$
in the full energy region $\sqrt{t}<700$ MeV, (where $t$ is the
square of the total energy in $\g\g$ c.m. system). 
The results obtained in these 
works are close to ChPT predictions \cite{gasser,burgi}. However, 
the values 
of the experimental cross section of the process $\ggc$ in this region
\cite{pluto,dm1,dm2,mark} are very ambiguous, 
and, as has been shown in Ref. \cite{holst,fil3},
even changes of these values by more than 100\% are still
compatible with the present error bars. 


Therefore, it is necessary to consider other additional possibilities of
the $\amc$ determination.

Such an information could be obtained from the dispersion
sum rules (DSR) for these parameters. However, the main contribution to
DSR for $\amc$ is given by the $\s$ meson, which is very wide and this
causes additional uncertainties in the DSR calculation.

On the other hand, if we consider the DSR for the difference between the charged
and neutral meson polarizabilities $\Delta(\am)=(\amc-\amn)$, then the 
contribution of mesons with isotopic spin $I=0$ in the $t$ channel to this 
difference would be equal 0 \cite{lvov1,fgr}, when the masses 
of charged and neutral $\pi$ mesons are equal each other. As a result, 
a model dependence for $\amc$ should be decreased essentially.

In the present work we investigate DSR for $\dcn$ as a function of the 
decay width of $\sigma\to\g\g$, when the masses of the charged and neutral 
$\pi$ mesons are not equal each other. As will be shown, the contribution 
of the $\sigma$ meson is small in this case too, and we can find a realistic 
limit on the value of $\amc$.

It has been shown that the value of the decay width of the $h_1(1170)$ meson 
into $\g\pi^0$ can be found if the value $\dcn$ is reliably determined
from experiment.


\section{Dispersion sum rules for the pion polarizabilities}

We will consider the helicity amplitudes $\mpp$ and $\mm$.
These amplitudes have no kinematical singularities or zeros \cite{aber}.
The relations between the amplitudes $\ggc$, $\gg0$ and the ones with 
isotopic spins $I=0$ and $I=2$ read
\beq
F_C&=&\sqrt{\frac{2}{3}}\left(F^0 +\frac{1}{\sqrt{2}} F^2\right), \nonumber \\
F_N&=&\sqrt{\frac{2}{3}}\left(F^0 -\sqrt{2} F^2\right).
\label{iso}
\eeq

The dipole ($\alpha_1$ and $\beta_1$)
polarizabilities are defined \cite{rad,fil2} through
expansion of the non-Born helicity amplitudes of Compton scattering on 
the pion in powers of $t$ at fixed $s=\m$
\beq
\mpp(s=\m,t)&=&2\pi\mu\am +{\cal O}(t),
\nonumber \\
\mm(s=\m,t)&=&2\frac{\pi}{\mu}\ap +{\cal O}(t),
\label{mpm}
\eeq
where $\m$ is the $\pi$ meson mass (different for $\pi^0$ and $\pi^{\pm}$),
$t+s+u=2\m$.

The dispersion sum rules for the difference of the dipole 
polarizabilities was obtained
in Ref. \cite{fil1} using DRs at fixed $u=\m$ without subtractions
for the amplitude $\mpp$. In this case, the Regge-pole model allows
the use of DR without subtractions \cite{aber}. Such a DSR is
\beq
\am&=&
\frac{1}{2\pi^2\mu}\left\{\int\limits_{4\m}^{\infty}~\frac{
Im\mpp(\tp,u=\m)~d\tp}{\tp}\right. \nonumber \\ 
& & \left.+\int\limits_{4\m}^{\infty}~\frac{
Im\mpp(\sp,u=\m)~d\sp}{\sp-\m}\right\}.
\label{dsr1m}
\eeq

As is evident from Eq.(\ref{iso}), the contribution of the isoscalar mesons
to the difference $\dcn$ equals 0 
(if the masses of the charged and neutral pions are
equal). We will study this difference when these masses do not equal
each other.

The DSRs for the charged pions are saturated  by the contributions of
the $\rho(770)$, $b_1(1235)$, $a_1(1260)$, and $a_2(1320)$ mesons in
the $s$-channel and $\s$, $f_0(980)$, $f_0^{\i}(1370)$ 
in the $t$-channel. For the $\pi^0$ meson the contribution of the $\rho$,
$\omega(782)$, $\phi$, $h_1(1170)$, and $b_1(1235)$  mesons are considered in 
the $s$-channel and the
same mesons as for the charged pions in the $t$-channel. Besides, we
take into account a nonresonant S-wave contribution of two charged
pions in the t channel. 

The parameters of the $\rho$, $\omega$, $\phi$, $b_1$ and $a_2$ mesons 
are given by the Particle Data Group \cite{pdg}. 
For the $a_1(1260)$ meson we took 
$m_{a_1}=1230$ MeV \cite{pdg}, $\G_{a_1}=$425 MeV (the average
value of the PDG estimate \cite{pdg}),
$\G_{a_1\to \g\pi^{\pm}}=0.64$ MeV \cite{zel}.

The parameters of the $f_0(980)$ and $f_0^{\i}(1370)$ mesons are taken as 
follows:

\noindent
$f_0(980)$: $m_{f_0}=980$ MeV \cite{pdg}, $\G_{f_0}=70$ MeV (the average
of the PDG \cite{pdg} estimate),
$\G_{f_0\to \g\g}=0.56\times 10^{-3}$ MeV,
$\G_{f_0\to \pi\pi}=0.84\,\G_{f_0}$ \cite{anis},
$\G_{f_0\to K\bar{K}}=0.16\,\G_{f_0}$;

\noindent
$f_0^{\i}(1370)$: $m_{f_0^{\i}(1370)}=1430$ MeV, $\G_{f_0^{\i}(1370)}=145$ MeV,
$\G_{f_0^{\i}(1370)\to \g\g}=0.54\times 10^{-5}$ MeV \cite{morg},
$\G_{f_0^{\i}(1370)\to \pi\pi}=0.26\,\G_{f_0^{\i}(1370)}$ \cite{bugg}.

The mass and the total decay width of the $h_1(1170)$ meson are taken
from PDG: $m_{h_1}=1170$ MeV, $\G_{h_1}=360$ MeV. The decay $h_1\to \g\pi^0$
has not yet been observed. Therefore we use this decay width according
the work \cite{garcia}:
\be
\G_{h_1\to\g\pi^0}=\frac{e^2}{4\pi}C_{h_1}\frac{(m_{h_1}^2-m_{\pi^0}^2)^3}
{3 m_{h_1}^3},
\ee
where the coefficient $C_{h_1}$ can be estimated using nonet symmetry
\cite{garcia,lvov2}:
\be
C_{h_1(1170)}\simeq 9 C_{b_1(1235)}\simeq 0.45.
\ee
As a result we have $\G_{h_1\to\g\pi^0}\simeq 1.6845\pm 0.44$ MeV.

Recently, a lot of works have been devoted to the study of the $\s$ meson
(see, for example \cite{capr1,capr2,mous,kamin,pela}). An average of
the most advanced data on the $\s$ meson gives (\cite{pela})
\be
m_{\s}=446\pm 6, \quad \Gamma_{\s}/2=276\pm 5.
\ee
In our analysis we use the values of the mass of the $\s$ meson and 
its total decay width in the following intervals:

$m_{\s}= 425 \div 550$ MeV, \quad $\Gamma_{\s}=400\div 600$ MeV.

The values of the decay width of $\s\to\g\g$ we consider from 0 up
to 3 keV.

Expressions for the imaginary parts of the resonances under
consideration are given in Appendix.

Besides the contribution of the $\s$, $f_0(980)$, and $f_0^{\i}(1370)$ mesons 
we have taken into account a nonresonant contribution of the $S$-waves with the 
isospin $I=0$ and 2 according to the diagrams of Fig. 1. 
\begin{figure}
\epsfxsize=8cm
\epsfysize=2cm    
\centerline{
\epsffile{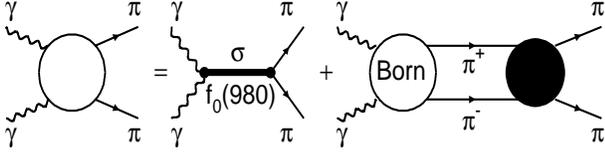}}              
\caption{$S$-wave contributions}
\end{figure}

It is worth noting that the vertexes of the $\s$ and $f_0$ meson poles in the 
dispersion approach include the full dynamics of the transitions on the 
mass shell. In this case there is no need to consider direct and rescattering 
mechanisms of transition separately.  

According to the unitarity condition,
the imaginary part of the amplitude $\mpp$ for the $\pi^+\pi^-$-loop diagram
in Fig. 1 can be written as
\be    
Im \mpp^{(s)}={\cal B}\, Re T_{\pi^+\pi^-\to \pi\pi},
\ee
where
${\cal B}\equiv {\cal B}(\ggc)$ is the contribution of the Born amplitude to the
$S$-wave of the $\ggc$ amplitude and equal to
\be \label{born}
{\cal B}=16\pi\left(\frac{e^2}{4\pi}\right)\frac{\mpi^2}{t^2}\ln\left(\frac{1+q/q_0}
{1-q/q_0}\right),
\ee
$q (q_0)$ is the momentum (energy) of the $\pi$ meson. The Born amplitude
can be expressed in terms of the $I=0$ and $I=2$ isospin amplitudes as
\be \label{bornI}
{\cal B}=\sqrt{\dfrac{2}{3}} B^{(I=0)}+\sqrt{\dfrac{1}{3}} B^{(I=2)}.
\ee
Taking into account that 
\be
{\cal B}(\gg0)=-\sqrt{\dfrac{1}{3}} B^{(I=0)}+\sqrt{\dfrac{2}{3}} B^{(I=2)}=0
\ee
we have \cite{bogl}
\be
B^{(I=0)}=\sqrt{\dfrac{2}{3}} {\cal B}, \quad B^{(I=2)}=\sqrt{\dfrac{1}{3}} 
{\cal B}.
\ee                          

The amplitudes of $\pi\pi$ scattering  are expressed through the amplitudes
in the isotopic space $T^{(I=0)}$ and $T^{(2)}$ as follows:
\beq \label{pipi}
T_{\pi^+\pi^-\to \pi^+\pi^-}&=&\frac{2}{3}\left(T^{(0)}+\frac{1}{2} 
T^{(2)}\right), \nonumber \\  
T_{\pi^+\pi^-\to \pi^0\pi^0}&=&\frac{2}{3}\left(T^{(0)}- T^{(2)}\right).
\eeq  

According to the relations (\ref{bornI}) and (\ref{pipi}) the imaginary parts 
of the $\pi\pi$ loop contributions to the $S$-wave of the amplitude $\mpp$ are
equal to
\be
Im \mpp^{(s)}(\gg0)=\frac{4}{9}{\cal B} Re\left(T^{(0)} + T^{(2)}\right),
\ee
\be
Im \mpp^{(s)}(\ggc)=\frac{1}{9}{\cal B} Re\left(4 T^{(0)} + T^{(2)}
\right).
\ee

The amplitudes $T^{(0)}$ and $T^{(2)}$ can be presented as
\be                       
Re\,T^{(I)}=\frac{q_0}{q}\eta_I \cos{\delta_0^I}(t) \sin{\delta_0^I}(t),
\ee
where $\delta_0^I(t)$ is the phase-shift of the $S$-wave of $\pi\pi$ scattering
with isospin $I$ and $\eta_I$ is the inelasticity.

The expression for the phase-shift $\delta_0(t)$ 
has been determined using 
the parameterization of Ref. \cite{garcia1}. At low energy $t\lesssim 4\mk$ 
we have
\beq
\cot{\delta_0^0(t)}&=&\frac{\sqrt{t}}{2 q} \frac{\m}{t-\frac{1}{2}\m}\left\{
\frac{\mu}{\sqrt{t}}+B_0+B_1 w(t) \right. \nonumber \\ 
 & & +\left.B_2 w(t)^2+B_3 w(t)^3 \right\},
\label{d01}
\eeq
where

$$ w(t)=\dfrac{\sqrt{t}-\sqrt{4\mk-t}}{\sqrt{t}+\sqrt{4\mk-t}},$$
and $\eta_0^0$ is equal to 1.  

For the energy $4\mk <t<$ (1.42 GeV)$^2$ we use \cite{garcia1}
\be
\delta_0^0(t) = d_0+B\frac{q_k^2}{\mk}+C\frac{q_k^4}{m_k^4} 
  +D\theta(t-4\me)\frac{q_\eta^2}{\me},
\label{d02}
\ee
\beq
\eta_0^0(t)&=&exp\left[\frac{-q_k}{\sqrt{t}}\left(\ep_1+\ep_2\frac{q_k}
{\sqrt{t}}+\ep_3\frac{q_k^2}{t}\right)^2\right. \nonumber \\
 & &\left.-\ep_4\theta(t-4\me)\frac{q_\eta}{\sqrt{t}}
\right],
\label{eta}
\eeq 
where $q_k=\sqrt{t/4-\mk}$ and $q_\eta=\sqrt{t/4-\me}$; $m_k$ and $m_{\eta}$
are the masses of $K$ and $\eta$ mesons, respectively. 

The parameters in Eqs(\ref{d01}-\ref{eta}) are listed in Table 1.

\begin{table}
\centering
\begin{tabular}{|c|l||c|l||c|l|}\hline
$B_0$ &$7.26\pm 0.23$ &$d_0$ &$(227.1 \pm 1.3)^{\circ}$ &$\ep_1$ 
&$4.7\pm 0.2$ \\ \hline
$B_1$ &$-25.3\pm 0.5$ &$B $ &$(94.0\pm 2.3)^{\circ}$ &$\ep_2$ 
&$-15.0\pm 0.8$ \\ \hline
$B_2$ &$-33.1\pm 1.2$ &$C $ &$(40.4\pm 2.9)^{\circ}$ &$\ep_3$ 
&$4.7\pm 2.6$ \\ \hline
$B_3$ &$-26.6\pm 2.3$ &$D $ &$(-86.9\pm 4.0)^{\circ}$ &$\ep_4$ 
&$0.38\pm 0.34$ \\ \hline
\end{tabular}
\caption{The values of the coefficients in Eqs. (\ref{d01},\ref{eta}).
\cite{garcia1}}
\end{table}

To describe the phase-shift $\delta_0^2(t)$ we use Schenk's 
parameterization \cite{schenk} in the energy
 region  up to 1.5 GeV, assuming that $\eta_0^2=1$ \cite{ananth}
\be
\tan{\delta_0^2}=\frac{q}{q_0}\left\{A0^2+B_0^2 q^2
 +C_0^2 q^4+D_0^2 q^6\right\}\left(\frac{4\m-s_0^2}{t-s_0^2}\right),
\ee
where
\beq
&& A_0^2=-0.044, \; B_0^2=-0.0855\m, \; C_0^2=-0.00754\mu^4,\nonumber \\  
&& D_0^2=0.000199\mu^6, \; s_0^2=-11.9\m .\nonumber 
\eeq

\section{Calculation of $\amc$}

The value of $\amn$ has been determined from the investigation of the
process $\gg0$ in the works \cite{fil1,kal,garcia}:
\noindent
$-1.6\pm 2.2$, $-0.6\pm 1.8$, $-1.25\pm 0.08\pm 0.15$.

These values are in good agreement with the prediction of ChPT \cite{bell}
$\amn^{ChPT}=-1.9\pm 0.2$. Therefore, in order to determine $\amc$ we
have added the value of $\amn^{ChPT}$ to the results of the calculations of
$\Delta(\am)$ with help of DSR (\ref{dsr1m}) at different values 
of the decay width of $\sigma\to\g\g$, when the mass and the total decay 
width of the $\s$ meson vary within the following values:
$\ms=400\div 550$ MeV, $\G_{\s}=400\div 600$ MeV.

The results of the calculation are shown in Fig.2. Line (1) 
corresponds to calculations with 
$\sqrt{t_{\s}}=\ms-i 1/2\G_{\s}=446-i 276$.
Lines (2) and (3)  correspond to $\ms=$400 MeV, $\G_{\s}=$600 MeV and
$\ms=$550 MeV, $\G_{\s}=$400 MeV, respectively. As is evident
from this Figure the values of $\amc$ weakly depend on the mass
and the total decay width of the $\s$ meson in the region under consideration.
The values of $\amc$ obtained are within $9.4\div 8.2$.
\begin{figure}
\epsfxsize=6cm
\epsfysize=8cm     
\centerline{
\epsffile{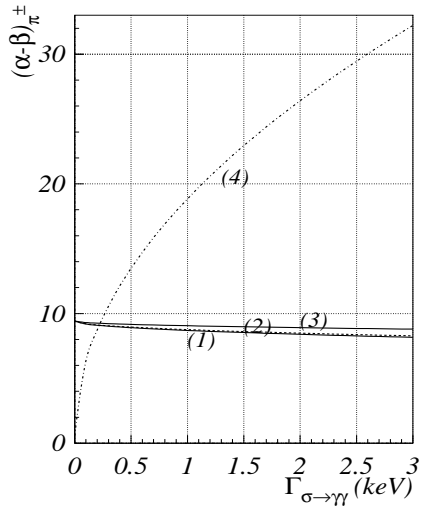}}              
\caption{Dependence of $\amc$ on $\G_{\s\to\g\g}$. Lines (1), (2), and (3)
correspond to the calculation of $\Delta(\am)$ at
$\sqrt{t_{\s}}=$446-i276, 400-i300, and 550-i200 MeV, 
respectively. Line (4) is the result of the calculations of DSR 
(\ref{dsr1m}) for $\amc$ at $\sqrt{t_{\s}}=$446-i276}   
\label{(a-b)}
\end{figure}

The greatest contribution to $\dcn$ is given by the $\omega$ and $h_1$
mesons. The parameters of the $\omega$ meson and so its contribution are well
known. On the other hand the experimental data on the $h_1$ meson are very
poor. In particular, the decay of this meson into $\g\pi^0$ was not
observed yet still in the experiment. Therefore, a reliable experimental
determination of $\dcn$ will allow to determine the real value of the
decay width $\G_{h_1\to\g\pi^0}$. For example, if the result of work
\cite{mainz} $\amc=11.6$ is confirmed, then 
$\G_{h_1\to\g\pi^0}=0.875$ MeV.

Line (4) in Fig.1 is the result of the calculations of DSR (\ref{dsr1m}) for
$\amc$ at $\ms=446$MeV and $\G_{\s}=552$MeV. This result strongly
depends on the decay width $\G_{\s\to\g\g}$ and indicates that 
realistic values of $\amc$ can be obtained if $\G_{\s\to\g\g}\lesssim 0.7$keV.

The influence of the upper integration limit ($\Lambda$) in the DSR 
(\ref{dsr1m}) on the results of the calculation was investigated. They are not
practically changed for $\Lambda$ more than (6 GeV)$^2$. In the present work we
performed the integration up to (20 GeV)$^2$.
                   
\section{Conclusions}

Using the fact that the contribution of the state with isospin $I=0$
to the difference $\Delta(\am)=\amc -\amn$ is very small
we have analyzed DSR for this difference
at the real values of the pion masses. DSR has been calculated for the $\s$ 
meson parameters within the intervals: $\ms=400\div 550$ MeV, 
$\G_{\s}=400\div 600$, $\G_{\s\to\g\g}=0\div 3$ keV. In order to
determine $\amc$ we have added $\amn^{ChPT}$=-1.9 to $\dcn$.
The values of $\amc$ found weakly depend on the $\s$ meson parameters 
and are in the range $\amc=9.4\div 8.2$. This result is in 
agreement with the experimental values obtained in work
\cite{mainz}, whereas it is at variance with the calculations in 
the framework of ChPT \cite{gasser}.

It has been shown that further experimental investigation of $\dcn$
can be an opportunity to determine the decay width $\G_{h_1\to\g\pi^0}$. 

Besides, the analysis of DSR for $\amc$ showed that more realistic
values of this parameter ($\amc< 15$) can be obtained with help of DSR 
(\ref{dsr1m}) if the decay width $\G_{\s\to\g\g}\lesssim 0.7$keV.
The values $\G_{\s\to\g\g}\lesssim 1$ keV were obtained early in works
\cite{fil1,fil2,dubn,achas} also. Results with $\G_{\s\to\g\g}>1$ keV
quoted in the resent literature are listed in \cite{mous,hofer}

\section*{Acknowledgements}

The author would like to thank H.J.~Arends, A.~Thomas, Th.~Walcher, 
V.L.~Kashevarov, and A.I. L'vov for useful discussions.
This work was supported by the Deutsche Forschungsgemeinschaft (SFB 443).

\renewcommand{\theequation}{A\arabic{equation}}
\setcounter{equation}{0}
\section*{Appendix A}
The contributions of the vector and axial-vector mesons $(\rho, \omega,
\phi, a_1$, and $b_1 )$ to $Im \mpp(s,u=\m)$ are calculated with the help 
of the expression
\be
Im\mpp^{(V)}(s,u=\m)=\mp 4g_V^2 s\frac{\G_0}{(m_V^2-s)^2+\G_0^2}
\ee
where $m_V$ is the meson mass, the sign "+" corresponds to the
 contribution of the $a_1$ and $b_1$ mesons and
\be
g_V^2=6\pi\sqrt{\frac{m_V^2}{s}}\left(\frac{m_V}{m_V^2-\m}\right)^3
\G_{V\to\g\pi} D_1(m_V^2)/D_1(s),
\ee
\be
\G_0=\left(\frac{q_i^2(s)}{q_i^2(m_V^2)}\right)^{\frac32}\frac{m_V^2}
{\sqrt{s}}D_1(m_V^2)/D_1(s)\G_V
\ee
Here $D_1$ is connected with the centrifugal potential and equal to
$D_1=1+(q_i r)^2$ \cite{blatt}, 
$r=1$fm is an effective interaction radius,
$\G_V$ and $\G_{V\to \g\pi}$ are the total decay width and the decay 
width into $\g\pi$ of these mesons. 
The momentums $q_i^2$ for $(\rho,\,\omega,\,\phi,\,a_1$, and $b_1 )$ mesons
are equal to $(s-4\m)/4$, $(s-9\m)/4$, $(s-4\mk)/4$,
$(s-(m_{\rho}+\m)/)4$, and $(s-16\m)/4$, respectively.
\renewcommand{\theequation}{B\arabic{equation}}
\setcounter{equation}{0}

\section*{Appendix B}

The amplitude of the contribution of a scalar meson 
to the process $\ggn$ can be written as
\be \label{tampl}
T=\frac{g_s}{\sqrt{t} -M_s-i\frac{1}{2}\G_{s}}.
\ee 
Then it is easy to show that the imaginary part of the amplitude 
$Im\mpp^{\s}(t)$ 
of the $\s$ meson contributions to the process under consideration could be
presented as
\be \label{Iampl}
Im\mpp^{\s}(t)=\dfrac{g_\s(\sqrt{t}+M_s)\G_0^{\s}(t)}{(t-M_{\s}^2)^2+
(\G_0^{\s}(t))^2},
\ee
where
\be \label{gampl}
g_\s=\frac{8\pi}{t}\left [\frac{2}{3} \dfrac{M_\s \G_{\g\g} \G_\s}
{\sqrt{M_{\s}^2-4\m}} \right ],                   
\ee

\be \label{g0ampl}
\G_0^\s=\dfrac{M_\s (\sqrt{t}+M_\s)}{2 \sqrt{t}} \left(\dfrac{t-4\m}
{M_\s^2-4\m} \right)^{1/2}\G_{\s} .
\ee
These expressions (\ref{Iampl})-(\ref{g0ampl}) can be very useful to describe
scaler mesons with large decay widths. 

As the two $K$ mesons give a big contribution to the decay width of the
$f_0(980)$ meson and the threshold of the reaction 
$\g\g\to K\overline{K}$ is very close to the mass of the $f_0(980)$ meson,
we consider Flatt{\'e}'s expression \cite{flatte} for the $f_0(980)$ meson 
contribution to the process $\ggn$.

For $t>4\mk$:
\be
Im \mpp^{f_0}=g_{f_0}\frac{\gf0}{(\mf^2-t)^2+\gf0^2},
\ee
where
\beq
\gf0 &=&\left[\gfp\left(\frac{t-4\m}{\mf^2-4\m}\right)^{1/2}
\right.
 \nonumber \\
&& \left.+\gfk\left(\frac{t-4\mk}{\mf^2-4\mk}\right)^{1/2}\right]\mf.
\eeq

For $t<4\mk$:
\beq
&&Im \mpp=g_{f_0} \gf0\left(\left[\mf^2-t \right.\right.  \nonumber \\
&&\left.\left.-\left(\dfrac{4\mk-t}
{\mf^2-4\mk}\right)^{1/2}\mf \gfk \right]^2 +\gf0^2 \right)^{-1}, 
\eeq
\be
\gf0=\gfp\mf \left(\frac{t-4\m}{\mf^2-4\m}\right)^{1/2}.
\ee

\end{document}